\documentclass[pra,twocolumn,a4paper,superscriptaddress]{revtex4}
\usepackage{amsmath}
\usepackage{amsfonts}
\usepackage{graphicx}
\begin{document}

\title{Non-classicality of photon added coherent and thermal radiations}
\author{A. R. Usha Devi}
\author{R. Prabhu}
\email{prabhurama@gmail.com}
\author{M. S. Uma}
\affiliation{Department of Physics, Bangalore University, Bangalore-560 
056, India}

\date{\today}

\begin{abstract}
Production and analysis of non-Gaussian radiation fields has evinced a lot of attention recently. 
Simplest way of generating such  non-Gaussians is through adding (subtracting) photons to 
Gaussian fields. 
Interestingly, when photons are added to classical Gaussian fields, the resulting states  exhibit 
{\em non-classicality}. 
Two important classical Gaussian radiation fields are coherent and thermal states. Here, we  
study the non-classical features of 
such states when photons are  added to them. Non-classicality of these states shows up in the 
negativity of the 
Wigner function.  We also work out the {\em entanglement potential}, a  recently proposed measure 
of non-classicality for these states. Our analysis reveals that photon added coherent states are 
non-classical for all seed beam intensities; their non-classicality increases with the addition 
of more number of photons. Thermal state exhibits non-classicality at all temperatures, when a 
photon is added; lower the temperature, higher is their non-classicality. 
\end{abstract}

\maketitle

\section{\label{sec:level1}Introduction}

Several branches of quantum optics from non-linear optics to laser-physics and cavity QED are 
very actively engaged in a variety of processes producing  non-classical light. Such radiation 
fields attract attention, not only because they provide a platform for testing fundamental 
concepts of quantum theory, but also for applications of importance like precision measurements 
in interferometry~\cite{martini}. Moreover, rapidly developing area of quantum computation and 
information theory has kindled further interest in generating and manipulating non-classical 
radiation fields - called quantum continuous variable states. These states are promising 
candidates for many applications of quantum information technology~\cite{tittle,zol}. In such a 
context, Gaussian light fields gain prominence, both in view of their conceptual and experimental 
importance. However, a need to leap beyond Gaussian domain has been emphasized~\cite{gran} and 
the degaussification process has been catching a lot of interest. Degaussification can be 
realized in a simple manner by adding (subtracting) photons to (from) a Gaussian field and the 
resulting states are known to exhibit non-classical properties such as negativity of the Wigner 
function~\cite{wf}, antibunching~\cite{kim}, sub-poissonian photon statistics~\cite{short} or 
squeezing in one of the quadratures of the field~\cite{dod1} etc.

Almost a decade ago  Agarwal and Tara~\cite{aga} introduced, theoretically, a new class of 
non-Gaussian states, which is obtained by repeated application of the photon creation operator on 
the  coherent state. The resulting class of states were identified to lie between the Fock state 
and the coherent state and were indeed non-classical. Recently, single photon excitation of a 
classical coherent field has been generated experimentally~\cite{belli} and ultrafast, 
time-domain, quantum homodyne tomography technique has explicitly demonstrated a quantum to 
classical  transition.  In another development, a traveling non-Gaussian field was experimentally 
produced by subtracting a photon from a squeezed vacuum~\cite{gran}. While the pulsed homodyne 
detection scheme confirmed non-Gaussian statistics for the photon subtracted squeezed vacuum, the 
Wigner function reconstructed from the experimental data failed to exhibit negativity. Kim {\em 
et al.}~\cite{mskim} analyzed the non-classicality of photon subtracted Gaussian fields and 
identified that the photo detection efficiency as well as the modal purity parameter were not 
high enough to record a negative Wigner function in the experiment~\cite{gran} of Wenger {\em et 
al.} Moreover, Kim {\em et al.} show that unless the input Gaussian radiation is non-classical, 
one cannot generate a non-classical field through photon subtraction~\cite{mskim}. They also 
point out the contrasting situation of photon addition to Gaussian fields, where even a highly 
classical state like thermal state turns out to be non-classical~\cite{ctlee,jones}. It may be 
noted that photon addition to a thermal state results in the removal of the vacuum 
part~\cite{jones} and hence leads to a {\em truncated thermal state}. All such {\em truncated 
states}, with their vacuum contribution removed, are shown to be non-classical by 
Lee~\cite{ctlee}.

In view of the current experimental progress~\cite{gran,belli} in the production of non-Gaussian 
radiation fields, and also in the tomographic reconstruction~\cite{gran,belli,lvovsky} of Wigner 
functions of quantum states, it is timely to analyze the non-classicality of photon added 
classical radiation fields, \,\, through negativity of their Wigner functions. In this paper, we 
investigate photon added coherent and thermal states. 

Entanglement, another striking quantum feature, has occupied a central position in the 
development of quantum information processing~\cite{tittle,zol}. There has been a considerable 
progress in understanding the connection between non-classicality and entanglement. It has been 
identified~\cite{ash,paris} that non-classicality is an unmistakable source of entanglement.
A beam splitter is capable of converting non-classicality of a single mode radiation into 
bipartite entanglement. This property viz., {\em non-classicality as an entanglement resource}, 
has been employed recently~\cite{ash}, to identify {\em Entanglement potential} (EP) - a 
computable measure of non-classicality - of single mode radiation fields. EP allows us to analyze 
the degree of non-classicality of a given single mode radiation field. We compute EP of photon 
added coherent states (PACS) and show that the EP reduces with the increase of seed beam 
intensity, which is in confirmation with the analysis of the Wigner function of the state. A 
comparison of EP's of single, two and three photon added coherent states reveals that 
non-classicality of PACS increases with the addition of more number of photons. We verify that 
the Wigner function of a photon added thermal state is negative at the phase space origin for all 
temperatures, while the EP of the state, evaluated in the low temperature limit, reveals that 
non-classicality of the state reduces with increasing temperature.

\section{\label{sec:level2} Measures of non-classicality}
Generally a non-classical state is recognized as one, which cannot be written as a statistical 
mixture of coherent states. It has been well accepted that the non-existence of a well defined 
Glauber-Sudarshan P-function~\cite{sudar} implies {\em non-classicality} of a given state. 
However this identification poses operational difficulties, as it requires complete information 
of the state to be examined, so that it's P-function can be reconstructed. Several operational 
criteria, which are equivalent to the one based on the P-function and which can be used to 
distinguish between classical and non-classical states in experimental measurements have been 
proposed from the early days of quantum optics. Such signatures of non-classicality, verifiable 
in a simple experiment are, antibunching~\cite{kim} and sub-poissonian photon 
statistics~\cite{short}, squeezing~\cite{dod1}, photon number oscillations~\cite{Schleich}, 
negative value of Wigner function~\cite{wf}, etc. 

Here, we focus our attention on the Wigner function of a given quantum state. 
The non-classicality character of a state is strongly registered by  negativity of the  Wigner 
function. Especially, Fock states show a negative dip around the phase space origin, as has been 
clearly reflected in the experimentally reconstructed Wigner function~\cite{lvovsky}. Moreover, 
there has been an ongoing effort towards more efficient quantum homodyne tomographic 
techniques~\cite{gran,belli,lvovsky}, and in such a context,  analysis of Wigner function proves 
to be useful. 

The Wigner function of a system, characterized by the density operator $\rho$ is 
defined~\cite{wf} through
\begin{equation}
\label{wigner}
W(q,p)=\frac{1}{\pi}\int\langle q+y|\rho |q-y\rangle \,e^{-2ipy}\, dy\,\,.
\end{equation}
Basically Wigner function is a quasi-probability distribution representing quantum  states  in 
phase space. It is not a true probability distribution as it  can take negative values 
also. If, for a state, Wigner function takes negative value, the quantum state has no classical 
analog. However, the converse does not hold good: When the Wigner function is positive 
everywhere, one can not conclude that the state is classical. For example, for a squeezed state, 
Wigner function is a Gaussian and is positive throughout. But, squeezed radiation~\cite{dod1} is 
one of the most important non-classical field. Thus, one has to resort to other measures of 
non-classicality.

There have been several approaches to quantify non-classicality of a state through universal 
measures like, \, Hillery's non-classical distance~\cite{hill} and Lee's non-classical 
depth~\cite{leedepth}. Distance between a given non-classical state and the set of all classical 
states is non-zero and hence, serves as a measure of non-classicality, called {\em non-classical 
distance}~\cite{hill}. However, identifying an optimal reference  classical state is one of the 
main problems associated with these distance based measures~\cite{manko} of non-classicality. 

 Lee's~\cite{leedepth} non-classical depth, $0\leq \tau_m \leq 1$, is essentially the quantity of  
smoothing required to transform  a non-positive P-function  into a well behaved positive 
distribution. (Non-classical depth $\tau_m$ is also interpreted as the minimum average number of 
thermal photons that are necessary to destroy the non-classical effects of a given state).  A 
classical state has $\tau_m=0.$
For a pure Gaussian state (Squeezed vacuum), the non-classical depth varies between 0 and 
 $\frac{1}{2}$~\cite{leedepth}; for all non-Gaussian pure states $\tau_m=1$~\cite{barn}, thus 
placing all such states to be 
 identical,  as far as their non-classicality is concerned.  Therefore, this criterion forces one 
to conclude that the non-classicality of PACS is independent of the seed beam intensity. 
Moreover, non-classical depth of photon added thermal states (for that matter, all {\em truncated 
states}) is shown to  be a maximum i.e.,  $\tau_m=1$~\cite{jones}.  According to this measure, 
photon added thermal states are equally (maximally) non-classical at all temperatures.

In this paper, we consider Entanglement potential (EP) ~\cite{ash}, as a universal measure of 
non-classicality for our discussion of PACS and photon added thermal states. EP gives  the amount 
of two-mode entanglement that can be  generated from a non-classical input state, in a linear 
optics set up. It is important to note here that a classical single mode radiation {\em does not} 
get entangled in such an arrangement~\cite{paris}. EP is nothing but the logarithmic 
negativity~\cite{vidal} of a bipartite quantum state $\varrho_{\sigma},$ which results from  
mixing a given single mode state $\sigma$, with vacuum state, in a $50:50$ beam splitter. More 
specifically,  EP is defined as
\begin{equation}
{\rm EP}={\rm log}_{\,2}\,\|\varrho_\sigma^{PT}\|_1,
\label{epot}
\end{equation}
where $\varrho_\sigma^{PT}$ denotes the  partial transpose of a two-mode density operator 
$\rho_\sigma=U_{BS}(\sigma \otimes|0\rangle \langle 0|)U_{BS}^\dagger$.
In equation (\ref{epot}), the symbol $\|\,\cdot\,\|_1$ denotes the trace norm\footnote{Trace norm 
of a partially transposed density operator $\varrho^{PT}$ is given by $\|\varrho^{PT}\|_1=1+2 
N(\rho)$, where the negativity $N(\rho)$ is the sum $|\sum_{i}\lambda_i|$ of all the negative 
eigenvalues of $\varrho^{PT}$.} and $U_{BS}$ corresponds to a $50:50$ beam splitter i.e.,  \break 
$~U_{BS}~=~{\rm exp}\,(\frac{\pi}{2}(a^\dagger b-a \,b^\dagger)),$   whose action (Heisenberg 
view point) on the creation operators $a^\dagger, b^\dagger$ of the two input ports is explicitly 
given by 
\begin{eqnarray}
U_{BS}\,\,a^\dagger\, U_{BS}=\frac{1}{\sqrt 2}\,(a^\dagger+b^\dagger)\nonumber \\
U_{BS}\,\,b^\dagger\, U_{BS}=\frac{1}{\sqrt 2}\,(b^\dagger-a^\dagger).
\label{unitary}
\end{eqnarray}
{\em The state $\sigma$ is said to be non-classical, iff its entanglement potential is nonzero}. 
EP has been  evaluated~\cite{ash} for a variety of 
non-classical states like squeezed states, even and odd coherent states, Fock states etc. 

The remainder of this paper is devoted to a study on photon addition to  
(i) coherent state,  an example of a pure Gaussian state and (ii) thermal state, a mixed Gaussian 
state, both of  which are well-known classical states. Action of photon creation operator on 
these states results in non-classical, non-Gaussian states.  We study the non-classicality of 
these states through negativity of the Wigner function and the entanglement potential.

\subsection{Photon added coherent state}

Coherent states are the analogs of classical radiation fields. These states are described by a 
Poissonian photon number 
distribution and have a well defined amplitude and phase. It is interesting to see how these 
states turn non-classical, when a 
single quantum of radiation excites them. 
\begin{figure}
 \includegraphics*[width=2.15in,keepaspectratio]{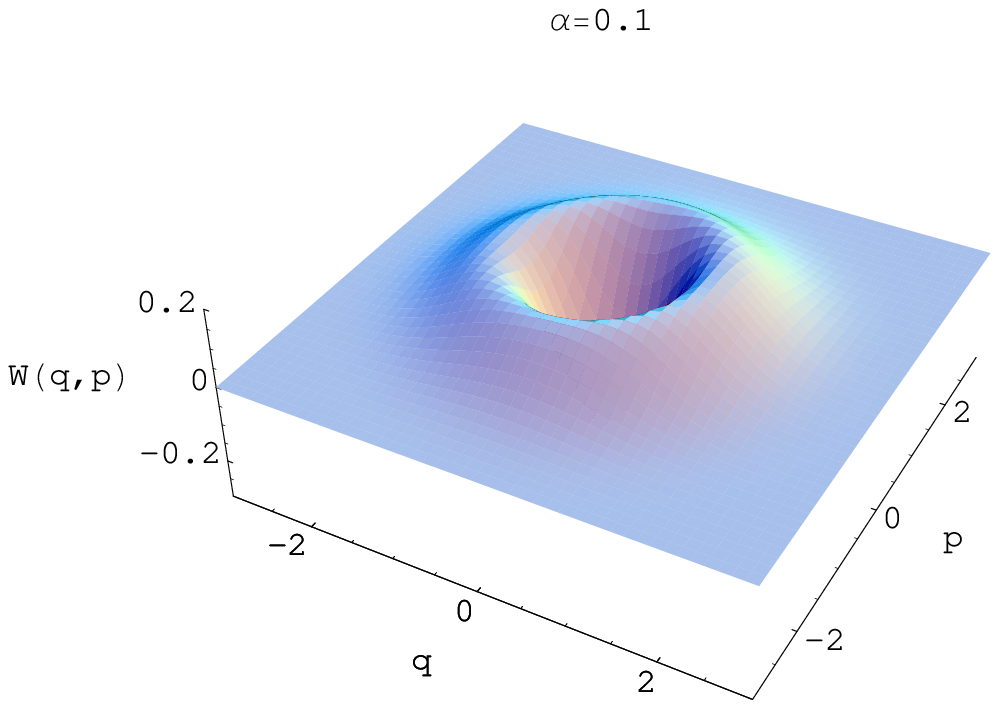} 
 \includegraphics*[width=2.15in,keepaspectratio]{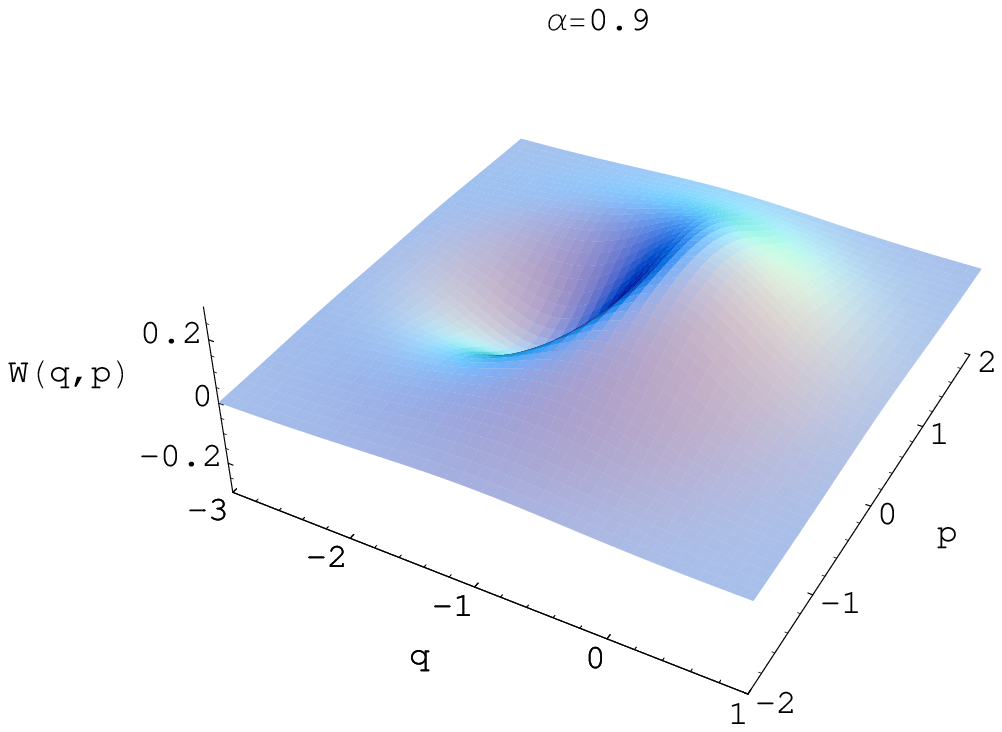}
 \includegraphics*[width=2.15in,keepaspectratio]{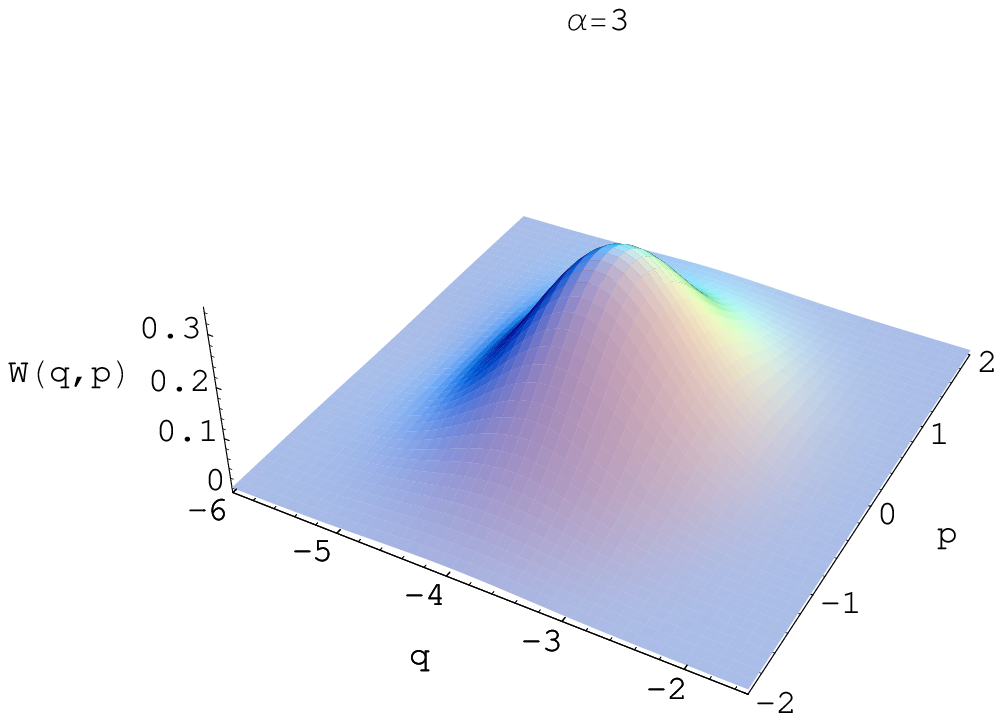}
 \caption{Wigner function of SPACS for different beam intensities $|\alpha|^2$. For simplicity, 
we have chosen $\alpha$ to be real. Here, (a) 
$\alpha=0.1$  (b) for $\alpha=0.9$  and (c) for $\alpha=3$.}
  \label{fig:spacs}
\end{figure}

Recently~\cite{belli}, single photon added coherent states\,\,\, [SPACS] has been generated
experimentally and tomographically reconstructed Wigner function for such states has been 
analyzed. SPACS are obtained by application 
of creation operator $a^\dagger$ on a coherent state $|\alpha\rangle$. Normalized SPACS is given 
by
\begin{eqnarray}
|{\rm SPACS}\rangle=\frac{a^\dagger|\alpha\rangle}{(1+|\alpha|^2)}&=&\frac{a^\dagger 
D_a(\alpha)\, |0_a\rangle}{\sqrt {(1+|\alpha|^2)}} \nonumber\\ \nonumber\\
&=&\frac{D_a(\alpha)\, [\, |1_a\rangle+\alpha^*|0_a\rangle]}{\sqrt {(1+|\alpha|^2)}},\,\,
\end{eqnarray}
where $D_a(\alpha)= {\rm exp}\, {(a^\dagger\alpha-a\alpha^*)}$ is the displacement operator and 
the coherent state $|\alpha\rangle=D_a(\alpha)|0_a\rangle$; $|0_a\rangle$ denotes the vacuum 
state. Here we have used the property,  $D_a^{\dagger}(\alpha)\,\,a^\dagger 
D_a(\alpha)=a^\dagger+\alpha^*$.  Wigner function of a SPACS 
is given by~\cite{aga}
\begin{eqnarray}
W(q,p)&=&\frac{-L_1(|2c-\alpha|^2)}{\pi L_1(-|\alpha|^2)}\,\,{\rm exp}{(-2|c-\alpha|^2)},
\end{eqnarray}
where $c=\frac{1}{\sqrt 2}(q+ip)\,\,\,{\rm and}\,\,\,L_1(z)=1-z$ is Laguerre polynomial of first 
order. Note that $W(q,p)$ of the SPACS is negative when $|2c-\alpha|^2<1$.

\begin{figure}
 \includegraphics*[width=2.15in,keepaspectratio]{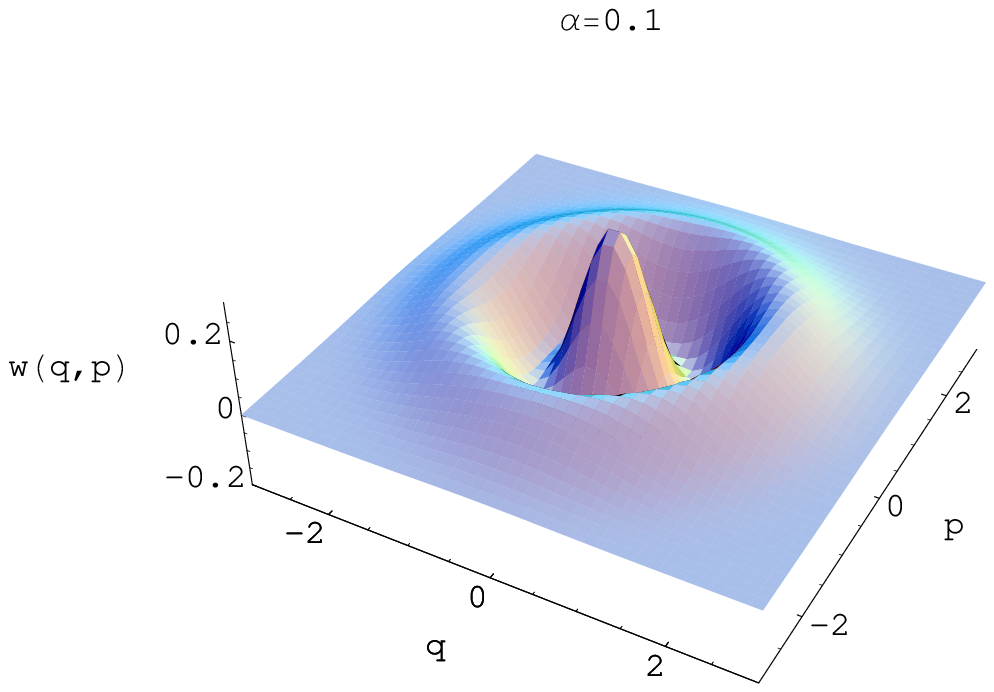} 
 \includegraphics*[width=2.15in,keepaspectratio]{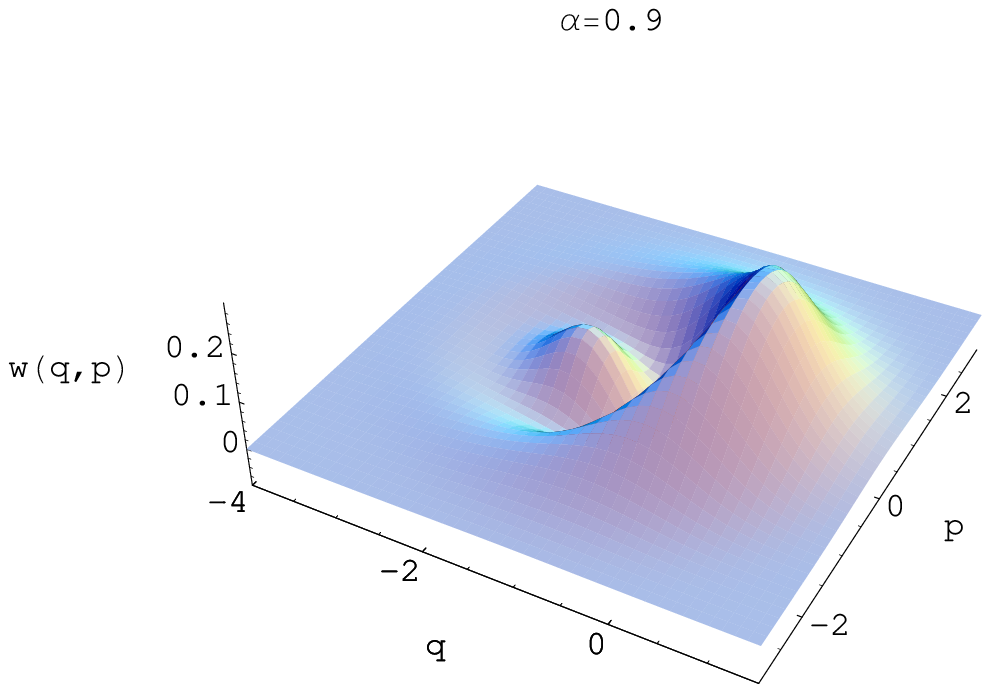}
 \includegraphics*[width=2.15in,keepaspectratio]{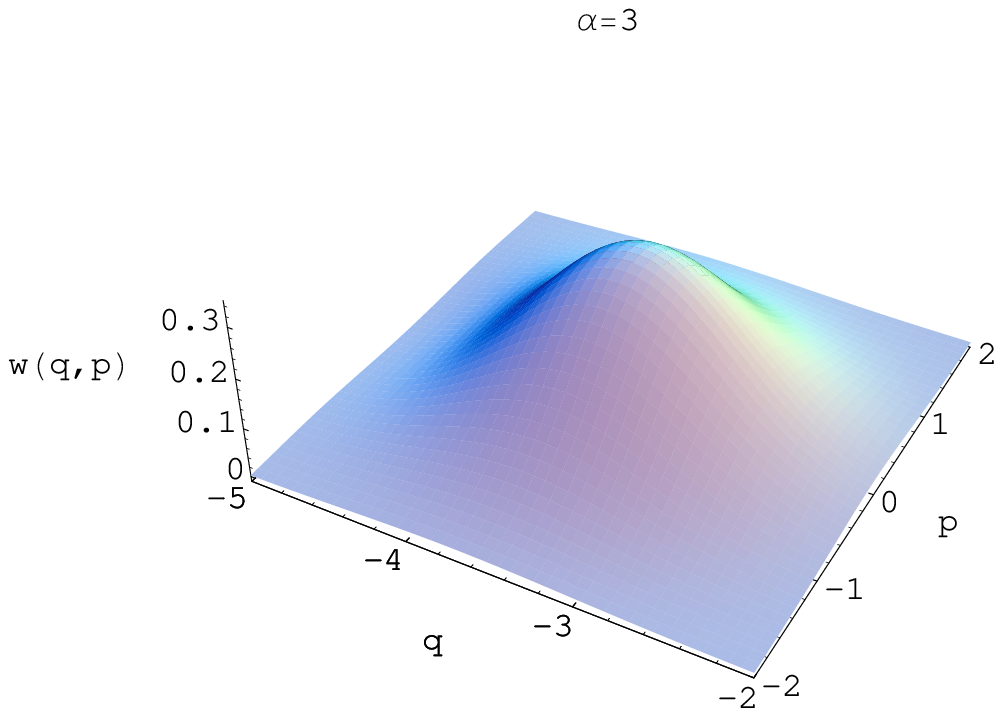}
  \caption{Wigner function of two PACS for different beam intensities $|\alpha|^2$. For 
simplicity, we have chosen $\alpha$ to be real. Here, (a) 
$\alpha=0.1$  (b) for $\alpha=0.9$  and (c) for $\alpha=3$.}
  \label{fig:tpacs}
\end{figure}
In Fig. \ref {fig:spacs} we have plotted the  Wigner function of SPACS, for various beam 
intensities $|\alpha|^2$. It is clear from these plots that negativity of Wigner function reduces 
with the increase of the intensity $|\alpha^2|$.  

We may verify whether addition of more number of photons leads to higher non-classicality of the 
beam. To see this we consider two photon added 
coherent states. Wigner function of such states is explicitly given by, 
\begin{eqnarray}
W(q,p)&=&\frac{L_2(|2c-\alpha|^2)}{\pi L_2(-|\alpha|^2)}\,\,{\rm exp}{(-2|c-\alpha|^2)},
\end{eqnarray}
where $L_2(z)=1-2z+\frac{1}{2}z^2$ is Laguerre polynomial of second order.

Fig. \ref {fig:tpacs}  gives the plots of the Wigner functions of two photon added coherent state 
for varying seed beam intensities. We note the same behavior here too, viz., the 
non-classicalities - depicted through the Wigner functions - decrease with increasing  intensity 
$|\alpha|^2$. But through these plots we can not conclude if addition of more number of photons 
leads to higher non-classicality or not. 

To investigate this, we now evaluate the EP of these photon added states. Let us first consider a 
SPACS. When SPACS is mixed with vacuum state $|0\rangle$ and is sent through a $50:50$ beam 
splitter, the resulting two mode state is given by
\begin{eqnarray}
|\psi\rangle=U_{BS}(|{\rm SPACS}\rangle \otimes|0\rangle)\hskip 2in\nonumber\\
=D_a\left(\frac{\alpha}{\sqrt 2}\right)D_b\left(\frac{\alpha}{\sqrt 
2}\right)\left[\frac{|1_a0_b\rangle+|0_a1_b\rangle+\sqrt 2 \, \alpha^*|0_a0_b\rangle}{\sqrt 
{2(1+|\alpha|^2)}}\right],\nonumber\\
\end{eqnarray}
since a $50:50$ beam splitter $U_{BS}$ acts on $D_a(\alpha)$ as (see equation (\ref{unitary}))
\begin{equation}
U_{BS}\,D_a(\alpha)\,U_{BS}^\dagger=D_a\left(\frac{\alpha}{\sqrt 
2}\right)D_b\left(\frac{\alpha}{\sqrt 2}\right),
\end{equation}
with $D_a(\frac{\alpha}{\sqrt 2})={\rm exp}\,({\frac{1}{\sqrt 2}(a^\dagger\alpha-a\,\alpha^*)})$ 
\,\,and\,\, $D_b(\frac{\alpha}{\sqrt 2})={\rm exp}\,({\frac{1}{\sqrt 
2}(b^\dagger\alpha-b\,\alpha^*)})$.   
The corresponding two mode density operator is given by
\begin{eqnarray}
\varrho'_0&=&|\psi\rangle\langle\psi|\nonumber\\
&=&D_a\left(\frac{\alpha}{\sqrt 2}\right)D_b\left(\frac{\alpha}{\sqrt 2}\right)\,\varrho _0\, 
D^\dag_a\left(\frac{\alpha}{\sqrt 2}\right)D^\dag_b\left(\frac{\alpha}{\sqrt 2}\right)\nonumber\\
\end{eqnarray}
with
\begin{equation}
\varrho_0=\frac{1}{2(1+|\alpha|^2)}
\left(
\begin{array}{cccccc}
2\,|\alpha|^2 & \sqrt 2\alpha^* & \sqrt 2\alpha^* & 0 & \cdot & \cdot \\
\sqrt 2\alpha & 1 & 1 & 0 & \cdot & \cdot \\
\sqrt 2\alpha & 1 & 1 & 0 & \cdot & \cdot \\
0 & 0 & 0 & 0 & \cdot & \cdot \\
\cdot & \cdot & \cdot & \cdot & \cdot & \cdot \\
\cdot & \cdot & \cdot & \cdot & \cdot & \cdot \\
\end{array}
\right)
\end{equation}
in the Fock state basis $|n_a \,\,m_b\rangle;\,\,\,\,\, n_a,\,m_b=0,1,2,\cdot\cdot\cdot$.
Since $\varrho'_0$ is locally equivalent to $\varrho_0 \,\,({\rm as}\, 
D_a\left(\frac{\alpha}{\sqrt 2}\right)D_b\left(\frac{\alpha}{\sqrt 2}\right)$ corresponds to a 
local displacements on the two mode states), EP of 
$\varrho'_0$ and that of $\varrho_0$ are same. So we proceed with the evaluation of EP of the 
state $\varrho_0$ itself. 

The partial transpose of $\varrho_0$ is given by\\ \\
\begin{equation}
\varrho^{\rm PT}_0=\frac{1}{2(1+|\alpha|^2)}
\left(
\begin{array}{ccccccc}
2\,|\alpha|^2 & \sqrt 2\alpha & \sqrt 2\alpha^* & 1 & 0 & \cdot & \cdot \\
\sqrt 2\alpha^* & 1 & 0 & 0 & 0 & \cdot & \cdot \\
\sqrt 2\alpha & 0 & 1 & 0 & 0 & \cdot & \cdot \\
1 & 0 & 0 & 0 & 0 & \cdot & \cdot \\
0 & 0 & 0 & 0 & 0 & \cdot & \cdot \\
\cdot & \cdot & \cdot & \cdot & \cdot & \cdot & \cdot \\
\cdot & \cdot & \cdot & \cdot & \cdot & \cdot & \cdot \\
\end{array}
\right),
\end{equation}
where, except for the first $4\times 4$ diagonal block all the other elements are zero. The 
non-zero eigenvalues of $\varrho^{\rm PT}_0$ are easily identified to be, 
\begin{eqnarray}
{\rm \lambda_1}&=&\frac{1}{2(1+|\alpha|^2)}, \ \ \ \ 
{\rm \lambda_2}=-\frac{1}{2(1+|\alpha|^2)}, \nonumber \\ \nonumber  \\
{\rm \lambda_{3,4}}&=&\frac{(1+|\alpha|^2)\pm\sqrt{(1+|\alpha|^2)^2-1})}{2(1+|\alpha|^2)}
\end{eqnarray} 
Note that $\lambda_2$ is the only negative eigenvalue of $\varrho^{\rm PT}$, and hence, the EP of 
a SPACS is given by,
\begin{eqnarray}
{\rm EP}&=&{\rm log_2}\|\varrho^{\rm PT}_0\|_1={\rm log}_2(1+2|\lambda_2|)\nonumber\\ 
&=&{\rm log_2\left(\frac{2+|\alpha|^2}{1+|\alpha|^2}\right)}.
\end{eqnarray}
\begin{figure}
  \includegraphics*[width=2.5in,keepaspectratio]{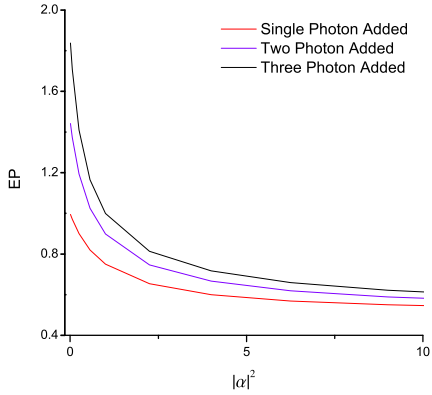}
  \caption{Entanglement potential for single, two and three photon added coherent states. (Here, 
$\alpha$ is chosen to be real for simplicity)}
  \label{fig:csln}
\end{figure}
Following similar lines, we can evaluate entanglement potentials of two and three photon added 
coherent states also. But the expressions are lengthy and do not exhibit a simple structure. 
We have computed them numerically and plots of entanglement potential for single, two and three 
photon added coherent states are given in Fig. \ref {fig:csln}. From the figure it is evident 
that EP is non zero for low intensity of the seed beam and it reduces gradually with the increase 
of intensity showing that the state is non-classical for all seed beam intensities. This 
observation is in confirmation with the conclusions reached through Wigner function analysis.  
While the EPs of  single, two and three photon added coherent states converge for higher beam 
intensities, they are all different for low beam intensity, with  larger value for higher photon 
added states. This observation reveals that non-classicality of photon added coherent state 
increases with the addition of larger number of photons.

\subsection{Photon added thermal state}

Density matrix of single mode thermal state of system in thermal equilibrium, characterized by 
the Hamiltonian $\hat H=a^\dagger a\, \hbar\omega$ is given by 
\begin{equation}
\rho_{\rm th}=\frac{{\rm exp}{\left(-\frac{\hat H}{kT}\right)}}{{\rm Tr}\left[{\rm 
exp}{\left(-\frac{\hat H}{kT}\right)}\right]}.
\end{equation}
In the Fock state basis, $\rho_{\rm th}$ can be expressed in the form 
\begin{equation}
\rho_{\rm th}=A\, \sum_{n=0}^\infty x^n \,|n\rangle\langle n|,
\end{equation}
where $A=1-x,\,\, x=e^{-\frac{\hbar\omega}{kT}}$;\ $0\leq x\leq 1$. Note that $x\rightarrow 0$ 
limit corresponds to $T\rightarrow 0$ and $x\rightarrow 1$ implies 
$T\rightarrow \infty.$
Photon added thermal state is obtained through the application of  creation operator on the  
thermal state i.e.,
\begin{eqnarray}
\label{th}
a^\dagger \rho_{\rm th}\, \, a&=&(1-x)\sum_{n=0}^\infty x^n \,a^\dagger\,|n\rangle\langle n|\, 
a\nonumber \\
&=&(1-x)\sum_{n=0}^\infty x^n \,(n+1)\,|n+1\rangle\langle n+1|\,.\nonumber\\ 
\end{eqnarray}
Simplifying equation (\ref {th}) we get, 
\begin{eqnarray}
a^\dagger \rho_{\rm th}\, \, a&=&(1-x)\sum_{m=0}^\infty m\,x^{m-1}\, |m\rangle\langle m| 
\nonumber\\
&=&(1-x)\frac{\partial }{\partial x}\left(\frac{1}{1-x}\,\rho_{\rm th}\right).
\end{eqnarray}
Normalized  photon added thermal state is given by  
\begin{eqnarray}
\rho_{\rm th}^{pa}&=&(1-x)^2\frac{\partial }{\partial x}\left(\frac{1}{1-x}\,\rho_{\rm 
th}\right).
\end{eqnarray}
Wigner function of a  thermal state has been identified~\cite{wf}  to be 
\begin{equation}
\label{with}
W_{\rm th}(q,p)=\frac{1}{\pi}\,B\,  {\rm exp}\, [-B\,(q^2+p^2)]; \, B=\frac{1-x}{1+x}. 
\end{equation}
\begin{figure}
 \includegraphics*[width=2.5in,keepaspectratio]{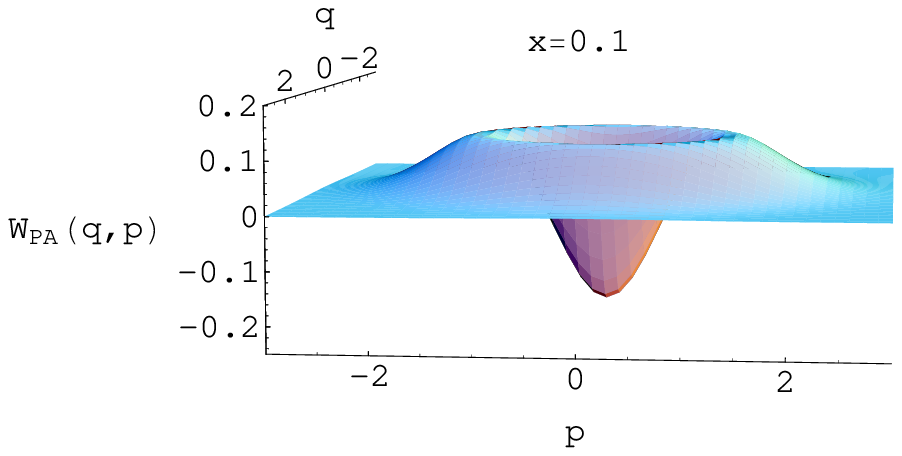} 
 \includegraphics*[width=2.5in,keepaspectratio]{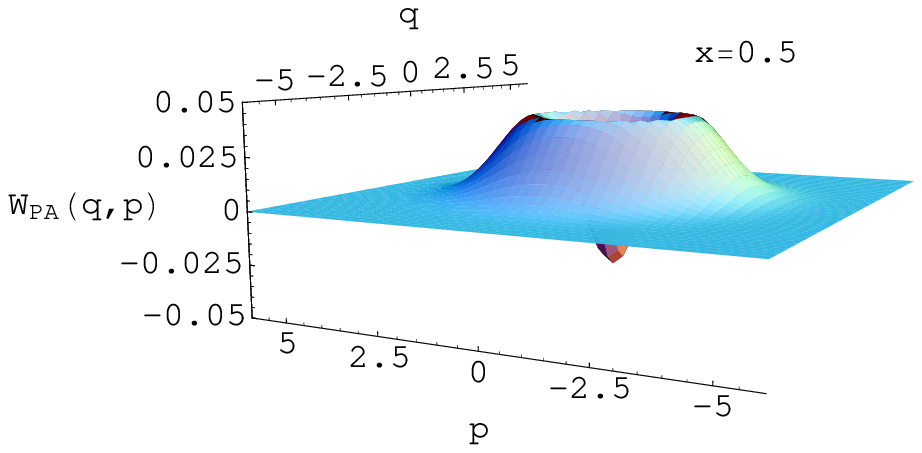}
 \includegraphics*[width=2.5in,keepaspectratio]{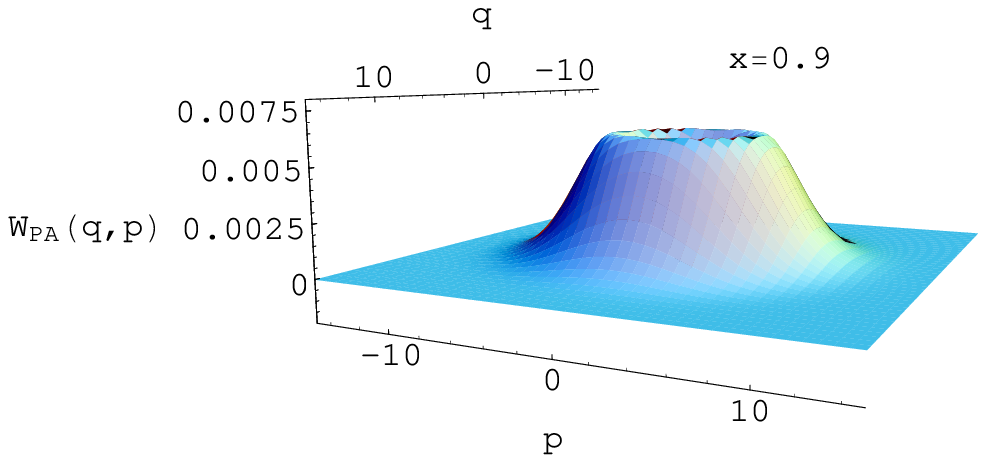}
\caption{Wigner functions of single photon added thermal state for various temperatures: (a) 
$x=0.1$ (b) $x=0.5$ \hskip 0.5 in (c) $x=0.9$.}
  \label{fig:tswf}
\end{figure}
Making use of the above equation, it is easy to evaluate the  Wigner function of a photon added 
thermal state:
\begin{equation}
W_{\rm th}^{pa}(q,p)=(1-x)^2\, \frac{\partial }{\partial x}\left(\frac{1}{1-x}\, W_{\rm 
th}(q,p)\right).
\end{equation}
After simplification, we get the Wigner function of the photon added thermal state as 
 \begin{equation}
W_{\rm th}^{pa}(q,p)=\frac{1}{\pi}\,B^2\left[\frac{2(q^2+p^2)}{(1+x)}-1\right]{\rm exp}\, 
[-B\,(q^2+p^2)]. 
\end{equation}
It is clear that the Wigner function $W_{th}^{pa}$ is negative at the origin of phase space, at 
all temperatures.
We have plotted  the Wigner function for different values of the parameter $x$ - which in turn 
corresponds to various temperatures - in Fig. \ref {fig:tswf}. It is clear from the plots that 
the photon added thermal states at various temperature are all non-classical. 

Entanglement potential of photon added thermal states can be evaluated in the low temperature 
limit, since the 
higher Fock states have lesser occupancy in this limit leading to the truncation of the Hilbert 
space.
Retaining terms up to first and  second order in the parameter $x$,  the density matrices of 
photon added thermal state are given below:  
\begin{eqnarray}
{\rm I\,\,order\, in\, }\, x: \,\,\, \rho _{th}^{pa}&=&(1-2x)\,\,|1\rangle\langle 
1|+2x\,|2\rangle\langle 2|\nonumber\\
{\rm II\,\,order\, in\, }\, x: \,\,\, \rho 
_{th}^{pa}&=&\frac{1}{(1-x^2)}\,[(1-2x)|1\rangle\langle 1|\nonumber\\
& &+2\,x\,(1-2x)|2\rangle\langle 2|+3\,x^2\,|3\rangle\langle 3|]\nonumber\\
\end{eqnarray}
Following similar steps to evaluate EP of SPACS, we can numerically compute the entanglement 
potentials of photon added thermal states too, in the 
 low temperature limit. With the help of this analysis we realize that EP reduces with increasing 
temperature. We therefore conclude that 
 non-classicality of photon added thermal states reduces  gradually with the increase of 
temperature.

In conclusion, we have analyzed the non-classicality of  photon added coherent states and thermal 
states, using (i) negativity of the Wigner function and (ii) entanglement potential. 
We have shown that  photon added coherent states are  non-classical for all seed beam 
intensities; the degree of non-classicality reduces with the increase of intensity. 
Photon added thermal states are shown to be  
non-classical at all temperatures and their non-classicality reduces with the increase of 
temperature.

We thank J. K. Asboth for useful comments and the Referees for their insightful suggestions in 
the light of which the paper has been revised.

\end{document}